\documentclass[12pt]{article}
\usepackage{epsfig,float,latexsym}

\date{}
\begin{document}

\newcommand{\beq}{\begin{equation}}
\newcommand{\eeq}{\end{equation}}
\newcommand{\nn}{\nonumber}
\newcommand{\bea}{\begin{eqnarray}}
\newcommand{\eea}{\end{eqnarray}}

\title{Braneworlds, Conformal Fields and the Gravitons}

\author{Rui Neves\footnote{E-mail: rneves@ualg.pt. Also at Centro Multidisciplinar de Astrof\'{\i}sica - CENTRA, Instituto Superior
  T\'ecnico, Avenida Rovisco Pais, 1049-001 Lisboa and Departamento de F\'{\i}sica, Faculdade de Ci\^encias e Tecnologia,
  Universidade do Algarve, Campus de Gambelas, 8005-139 Faro, Portugal.}\\
{\small \it Centro de Electr\'onica, Optoelectr\'onica e Telecomunica\c {c}\~oes (CEOT)}\\
{\small \it Faculdade de Ci\^encias e Tecnologia}\\{\small \it Universidade do
  Algarve}\\
{\small \it Campus de Gambelas, 8005-139 Faro, Portugal}}

\maketitle

\begin{abstract}
We investigate the dynamics of Randall-Sundrum $\rm{AdS}_5$ braneworlds with 
5-dimensional conformal matter fields. In the scenario with a 
compact fifth dimension the class of conformal fields with weight -4 is 
associated with exact 5-dimensional warped geometries which are stable under radion 
field perturbations and describe on the brane the dynamics of inhomogeneous dust, generalized dark radiation
and homogeneous polytropic dark energy. We analyse the graviton mode
flutuations around this class of background solutions and determine
their mass eigenvalues and wavefunctions from a Sturm-Liouville
problem. We show that the localization of gravity is not sharp enough
for large mass hierarchies to be generated. We also discuss
the physical bounds imposed by experiments in particle
physics, in astrophysics and in precise measurements of the low energy gravitational interaction.

\end{abstract}

\section{Introduction}
In the $\rm{AdS}_5$ Randall-Sundrum (RS) scenario \cite{RS1,RS2} 
our visible 4-dimensional (4D) Universe is a 3-brane world embedded
in a $Z_2$ symmetric 5D anti-de Sitter (AdS) space. In the RS1 model
\cite{RS1} the fifth dimension is compact and there are two 3-brane
boundaries. In this setting gravity is exponentially localized near the
hidden positive tension brane and decays towards the observable
negative tension brane. The hierarchy problem is then  
reformulated as an
exponential hierarchy between the weak and Planck scales \cite{RS1}. 
In the RS2 model \cite{RS2} the $\rm{AdS}_5$ orbifold has an infinite fifth
dimension and a single visible positive tension
brane to which the gravitational field is exponentially bound. 

In the visible brane the low energy theory of gravity is 
4D general relativity and the cosmology may be 
Friedmann-Robertson-Walker \cite{RS1}-\cite{TM}. 
In the RS1 model this requires the stabilization of 
the radion mode with for example a 5D scalar
field \cite{GW,WFGK,CGRT,TM}. Using the
Gauss-Codazzi formulation \cite{BC,SMS} many other braneworld
solutions have been discovered although a number of them have not
yet been associated with exact 5D spacetimes \cite{COSp}-\cite{RC1}.

In this paper we continue the research about the dynamics of a
spherically symmetric RS 3-brane when conformal matter fields
propagate in the bulk \cite{RC2}-\cite{RC4} (see also \cite{EONO}).
Some time ago \cite{RC2,RC3} we have discovered a new class of exact 5D dynamical
warped solutions which is associated with conformal fields
characterized by an energy-momentum tensor of
weight -4. These
solutions were shown to describe on the brane the dynamics of
inhomogeneous dust, generalized dark radiation and homogeneous
polytropic matter \cite{RC2,RC3}. The latter in particular refers to a
perfect dark energy fluid and describes the accelerated expansion of
our Universe. The radion may be
stabilized by a sector of the conformal bulk fields of weight -4 while
another sector generates the dynamics on the brane. The
stabilization requires a bulk fluid sector with a constant negative 5D
pressure and involves new warp functions \cite{RC4}. If the theory of gravity on the
brane deviates from that of Einstein
the existence of such dynamical geometries requires the presence of
non-conformal matter fields confined to the brane \cite{ER}. In this
work we analyse the graviton field perturbations around this class of 
background geometries and determine
their mass eigenvalues and wavefunctions from a Sturm-Liouville
problem. We show that gravity is not suficiently localized near the
positive tension branes to be able to generate large mass hierarchies. 
We also discuss
the physical bounds imposed by experiments in particle
physics, in astrophysics and in precise measurements of Newton's law
of gravity. 

\section{5D Einstein equations and conformal fields}

The most general non-factorizable
dynamical metric 
consistent with the $Z_2$ symmetry in the fifth dimension and
with 4D spherical symmetry on the brane is given by 
\beq
{\rm d}{\tilde{s}_5^2}={\Omega^2}\left({\rm d}{z^2}-{{\rm e}^{2A}}{\rm
    d}{t^2}+{{\rm e}^{2B}}{\rm d}{r^2}+{R^2}{\rm d}{\Omega_2^2}\right),\label{gm1}
\eeq
where $(t,r,\theta,\phi,z)$ are the coordinates mapping the
$\rm{AdS}_5$ orbifold. In this set $z$ is related to the cartesian
coordinate $y$ by
$z=l{{\rm e}^{y/l}}$, $y>0$, with $l$ the AdS radius. The functions $\Omega=\Omega(t,r,z)$, $A=A(t,r,z)$, $B=B(t,r,z)$ and
$R=R(t,r,z)$ are $Z_2$ symmetric. $R(t,r,z)$ represents the
physical radius of the 2-spheres and $\Omega$ is the warp factor
characterizing a global conformal transformation on the metric.

In the RS1 model the classical dynamics is defined by the 5D
Einstein equations,
\beq
{\tilde{G}_\mu^\nu}=-{\kappa_5^2}\left\{{\Lambda_{\rm B}}{\delta_\mu^\nu}+
{1\over{\sqrt{\tilde{g}_{55}}}}\left[\lambda\delta
\left(z-{z_0}\right)+\lambda'\delta\left(z-{{z'}_0}\right)\right]\left({\delta_\mu^\nu}-{\delta_5^\nu}
{\delta_\mu^5}\right)-{\tilde{\mathcal{T}}_\mu^\nu}\right\},
\label{5DEeq}
\eeq
where $\Lambda_{\rm B}$ is the negative bulk cosmological constant and
${\kappa_5^2}=8\pi/{{\rm M}_5^3}$ with ${\rm M}_5$ the fundamental 5D
mass scale. $\lambda,\lambda'$ are the brane tensions and ${\tilde{\mathcal{T}}_\mu^\nu}$ is
the stress-energy tensor of the matter fields which in 5D is
conserved, 

\beq
{\tilde{\nabla}_\mu}{\tilde{\mathcal{T}}_\nu^\mu}=0\label{5Dceq}.
\eeq

For a general 5D metric $\tilde{g}_{\mu\nu}$ (\ref{5DEeq}) and
(\ref{5Dceq}) form a complex system of differential equations. 
To solve it let us introduce some simplifying assumptions \cite{RC4}. First consider that the bulk matter is
described by conformal fields with weight
$s$. Under the conformal transformation
${\tilde{g}_{\mu\nu}}={\Omega^2}{g_{\mu\nu}}$ the stress-energy tensor
satisfies
${\tilde{\mathcal{T}}_\mu^\nu}={\Omega^{s+2}}{\mathcal{T}_\mu^\nu}$. Then separate the conformal tensor $\tilde{\mathcal{T}}_\mu^\nu$ in two sectors
$\tilde{T}_\mu^\nu$ and $\tilde{U}_\mu^\nu$ with the same weight $s$,
${\tilde{\mathcal{T}}_\mu^\nu}={\tilde{T}_\mu^\nu}+{\tilde{U}_\mu^\nu}$
where ${\tilde{T}_\mu^\nu}={\Omega^{s+2}}{T_\mu^\nu}$ and
${\tilde{U}_\mu^\nu}={\Omega^{s+2}}{U_\mu^\nu}$, and take $s=-4$. Finally consider $A=A(t,r)$, $B=B(t,r)$, $R=R(t,r)$ and
$\Omega=\Omega(z)$. Then (\ref{5DEeq}) leads to \cite{RC4} 
\beq
{G_a^b}={\kappa_5^2}{T_a^b},\label{4DECeq}
\eeq
\beq
{G_5^5}={\kappa_5^2}{T_5^5},\label{5DEeqz}
\eeq
\beq
6{\Omega^{-2}}{{({\partial_z}\Omega)}^2}+{\kappa_5^2}{\Omega^2}{\Lambda_{\rm
    B}}={\kappa_5^2}{U_5^5},\label{rswf1}
\eeq
\beq
\left\{3{\Omega^{-1}}{\partial_z^2}\Omega+{\kappa_5^2}{\Omega^2}
\left\{{\Lambda_{\rm B}}+{\Omega^{-1}}\left[\lambda\delta(z-{z_0})+\lambda'\delta(z-{{z'}_0})\right]\right\}\right\}{\delta_a^b}={\kappa_5^2}{U_a^b},\label{rswf2}
\eeq
where the latin indices represent the 4D coordinates
$t$, $r$, $\theta$ and $\phi$. On the other hand from (\ref{5Dceq}) we
also obtain \cite{RC4}
\beq
{\nabla_a}{T_b^a}=0\label{5DceqT1}
\eeq
and the equations of state $2{T_5^5}={T_a^a},\quad 2{U_5^5}={U_a^a}$.
$U_\mu^\nu$ turns out to be a constant diagonal tensor,
${U_\mu^\nu}=\mbox{diag}(-\bar{\rho},{\bar{p}_{\rm r}},{\bar{p}_{\rm
      T}},{\bar{p}_{\rm T}},{\bar{p}_5})$,
with the density $\bar{\rho}$ and pressures $\bar{p}_{\rm r}$,
$\bar{p}_{\rm T}$, ${\bar{p}_5}$ satisfying 
${\bar{p}_5}=-2\bar{\rho}=2{\bar{p}_{\rm r}}=2{\bar{p}_{\rm T}}$. Note
that ${\nabla_a}{U_b^a}=0$ is an identity. If ${T_\mu^\nu}=\mbox{diag}\left(-\rho,{p_{\rm r}},{p_{\rm T}},{p_{\rm T}},{p_5}\right)$ where
$\rho$, $p_{\rm r}$, $p_{\rm T}$ and $p_5$ are, respectively, the density and
pressures then its equation of state is re-written as
\beq
\rho-{p_{\rm r}}-2{p_{\rm T}}+2{p_5}=0.\label{eqst3}
\eeq

\section{Exact 5D warped solutions}

The $\rm{AdS}_5$ braneworld dynamics is defined by the solutions of 
(\ref{4DECeq}) to (\ref{eqst3}) \cite{RC4}. Solving (\ref{rswf1}) and (\ref{rswf2}) we obtain 
\beq
\Omega(y)={{\rm e}^{-|y|/l}}\left(1+{p_{\rm
      B}^5}{{\rm e}^{2|y|/l}}\right),\label{wfp5y}\;
\lambda={\lambda_{\rm RS}}{{1-{p_{\rm B}^5}}\over{1+{p_{\rm
        B}^5}}},\; \lambda'=-{\lambda_{\rm RS}}{{1-{p_{\rm
        B}^5}\exp(2\pi{r_{\rm c}}/l)}\over{1+{p_{\rm
        B}^5}\exp(2\pi{r_{\rm c}}/l)}},\label{wft4}
\eeq
where ${p_{\rm B}^5}={\bar{p}_5}/(4{\Lambda_{\rm B}})$,
${\lambda_{\rm RS}}=6/(l{\kappa_5^2})$ and $r_{\rm c}$ is the RS compactification scale. 

To determine the dynamics on the brane we solve (\ref{4DECeq}) and
(\ref{5DEeqz}) when $T_\mu^\nu$ 
satisfies (\ref{5DceqT1}) and (\ref{eqst3}) \cite{RC4}. Note that as long as
$p_5$ balances $\rho, {p_{\rm
    r}}$ and $p_{\rm T}$ according to (\ref{5DEeqz}) and (\ref{eqst3}), the 4D equation of state is not constrained. Three
examples corresponding to inhomogeneous dust, generalized dark
radiation and homogeneous polytropic matter were considered in
\cite{RC2} and \cite{RC3}. The latter describes the
dynamics on the brane of dark energy in the form of a polytropic
fluid. The 5D dark energy polytropic solutions are \cite{RC3}
\beq
{\rm d}{\tilde{s}_5^2}={\Omega^2}\left[-{\rm d}{t^2}+{S^2}
\left({{{\rm d}{r^2}}\over{1-k{r^2}}}+{r^2}{\rm
    d}{\Omega_2^2}\right)\right]+{\rm d}{y^2},
\label{dmsol1}
\eeq
where $S$ satisfies $S{\dot{S}^2}={\kappa_5^2}{{\left(\eta{S^{3-3\alpha}}+
a\right)}^{1\over{1-\alpha}}}/3-k S$. $\alpha$ and $\eta$ characterize
different polytropic phases. For $-1\leq\alpha<0$ the fluid is in
its generalized Chaplygin phase. 

\section{Radion stability}

\begin{figure}[H]
   \center{\psfig{file=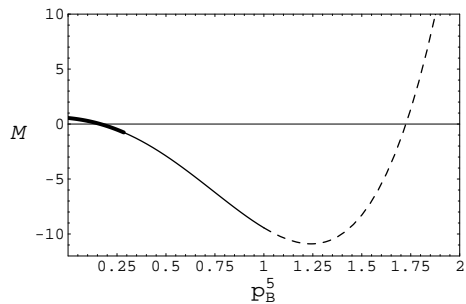,width=0.42\hsize}}
   \vspace{-0.2cm}
   \caption{Plot of radion mass parameter $M$ for
   $l/{r_{\rm c}}=5$. Thick line, $0<{p_{\rm B}^5}
     \leq{{\rm e}^{-2\pi/5}}: \lambda>0, \lambda'\leq 0$. Thin line,
     ${{\rm e}^{-2\pi/5}}
     <{p_{\rm B}^5}\leq 1: \lambda\geq 0,\lambda'>0$. Dashed line,
     ${p_{\rm B}^5}>1: 
     \lambda<0, \lambda'>0$.}
   \label{fig1:sta123}
   \end{figure}

Provided the equation of state of the conformal fluid is independent of the
radion perturbation, these solutions are stable for a range of the
model parameters if ${p_{\rm B}^5}>0$ \cite{RC4}. Using $x=y/{r_{\rm c}}$ the radion
mass is given in terms of the dimensionless radion mass parameter $M$ (see figure~\ref{fig1:sta123})
\beq
{m_\gamma}={1\over{r_{\rm c}}}\sqrt{{4|M|}\over{3{\int_{-\pi}^\pi}{\rm d}x{\Omega^2}}},\quad
M=\lambda{r_{\rm c}}{\kappa_5^2}{\Omega^4}(0)+\lambda'{r_{\rm c}}{\kappa_5^2}
      {\Omega^4}(\pi)-{{6{r_{\rm c}^2}}\over{l^2}}{\int_{-\pi}^\pi}
      {\rm d}x{\Omega^4}.
\eeq
For ${p_{\rm B}^5}>0$ the stability of the $\rm{AdS}_5$ braneworlds
also depends on the dimensionless ratio $l/{r_{\rm c}}$. For $l/{r_{\rm
         c}}<1.589\cdots$ all solutions turn out to be
     unstable. Stable universes begin to appear at $l/{r_{\rm
         c}}=1.589\cdots, {p_{\rm B}^5}=0.138\cdots$. For $l/{r_{\rm
         c}}>1.589\cdots$ we find stable solutions for an interval of 
$p_{\rm B}^5$ (see in figure~\ref{fig1:sta123} the example of $l/{r_{\rm
         c}}=5$) 
which increases with $l/{r_{\rm c}}$. For large
         enough but finite $l/{r_{\rm c}}$ the stability interval
         approaches the limit
         $\left]0.267\cdots,3.731\cdots\right[$. Naturally, $M\to 0$
         if $l/{r_{\rm c}}\to\infty$.

The particle physics bound for the radion mass is ${m_\gamma}>35-120$
GeV \cite{PDG}. For fixed $l/{r_{\rm c}}$ the order of magnitude bound
  ${m_\gamma}>$ 100 GeV requires a small 
fifth dimension, ${r_{\rm c}}<1\times{10^{-15}}$ mm. Since the 5D mass scale
is related to the 4D Planck mass ${{\rm M}_{\rm
    P}^2}=8\pi/{\kappa_4^2}\sim 1\times{10^{19}}$GeV by ${{\rm
    M}_5^3}{r_{\rm c}}{\int_{-\pi}^\pi}{\rm d}x{\Omega^2}={{\rm M}_{\rm P}^2}$ we conclude that ${{\rm M}_5}>2\times{10^{10}}$ TeV. Consequently this
radion is invisible in astrophysical processes which require at least
${m_\gamma}>$ 100 MeV \cite{PDG}. It is also invisible in precision Newton's
   law measurements which imply ${m_\gamma}>6.25\times{10^{-4}}$ eV \cite{PDG}.

If the scale ${\rm M}_5$ is high enough then the radion
   may be light. The direct bound from particle physics is ${{\rm M}_5}>1-2$
   TeV \cite{PDG}. From astrophysics it may be as high as
   ${{\rm M}_5}>1-200$ TeV \cite{PDG}. Now ${m_\gamma}>6.25\times{10^{-4}}$ eV
   corresponds to a large fifth dimension with ${r_{\rm c}}<160\mu$m. Then
   the 5D scale is indeed high, ${{\rm M}_5}>4\times{10^{5}}$TeV.

These results are a consequence of the absence of a
    large mass hierarchy, $m=\Omega(\pi{r_{\rm
    c}}){m_0},\quad\Omega(\pi{r_{\rm c}})\sim
   1-100$, and as order of magnitude estimates they hold for all the
    $p_{\rm B}^5$ stability range. 

\section{The gravitons}

To analyse the graviton field perturbations around the background
$\bar{g}_{\mu\nu}$ let us write the classical
5D Einstein equations as 
\beq
{\tilde{G}_{\mu\nu}}=-{\kappa_5^2}{\tilde{\mathcal{M}}_{\mu\nu}},
\eeq
where the energy-momentum tensor ${\tilde{\mathcal{M}}_{\mu\nu}}$ is
\[
{\tilde{\mathcal{M}}_{\mu\nu}}={1\over{\sqrt{\tilde{g}_{55}}}}\left[\lambda\delta
      \left(y-{y_0}\right)+\lambda'\delta\left(y-{{y'}_0}\right)\right]
      \left({\tilde{g}_{\mu\nu}}-{\delta_\mu^5}{\tilde{g}_{5\nu}}-{\tilde{g}_{\mu5}}{\delta_\nu^5}+{\tilde{g}_{55}}{\delta_\mu^5}{\delta_\nu^5}\right)
\]
\beq
+{\Lambda_{\rm B}}{\tilde{g}_{\mu\nu}}-{\tilde{\mathcal{T}}_{\mu\nu}}.
\eeq
Consider the graviton perturbation $h_{\mu\nu}$ such that
${\tilde{g}_{\mu\nu}}={\bar{g}_{\mu\nu}}+{h_{\mu\nu}}$. Using
${\hat{h}_{\mu\nu}}={h_{\mu\nu}}-{1\over{2}}{\bar{g}_{\mu\nu}}h$ where
$h={h_\mu^\mu}$ we find
\[
{\tilde{G}_{\mu\nu}}={\bar{G}_{\mu\nu}}-{1\over{2}}\left({\bar{\nabla}^2}{\hat{h}_{\mu\nu}}+{\bar{g}_{\mu\nu}}{\bar{\nabla}^\alpha}{\bar{\nabla}^\beta}{\hat{h}_{\alpha\beta}}\right)+{\bar{\nabla}_{(\mu}}{\bar{\nabla}^\alpha}{\hat{h}_{\alpha\nu)}}+{\bar{R}^{\alpha\;\beta}_{\,\mu\nu}}{\hat{h}_{\alpha\beta}}
\]
\beq
+{\bar{G}^\beta_{(\mu}}{\hat{h}_{\nu)\beta}}+{1\over{6}}\bar{R}\hat{h}{\bar{g}_{\mu\nu}}.
\eeq
On the other hand
${\tilde{\mathcal{M}}_{\mu\nu}}={\bar{\mathcal{M}}_{\mu\nu}}+\delta{\mathcal{M}_{\mu\nu}}$.
Since ${\bar{G}_{\mu\nu}}=-{\kappa_5^2}{\bar{\mathcal{M}}_{\mu\nu}}$
     we obtain
\[
-{1\over{2}}\left({\bar{\nabla}^2}{\hat{h}_{\mu\nu}}+{\bar{g}_{\mu\nu}}{\bar{\nabla}^\alpha}{\bar{\nabla}^\beta}{\hat{h}_{\alpha\beta}}\right)+{\bar{\nabla}_{(\mu}}{\bar{\nabla}^\alpha}{\hat{h}_{\alpha\nu)}}+{\bar{R}^{\alpha\;\beta}_{\,\mu\nu}}{\hat{h}_{\alpha\beta}}=
\]
\beq
-{\bar{G}^\beta_{(\mu}}{\hat{h}_{\nu)\beta}}-{1\over{6}}\bar{R}\hat{h}{\bar{g}_{\mu\nu}}-{\kappa_5^2}\delta{\mathcal{M}_{\mu\nu}}.
\eeq

Assume a flat brane background ${\bar{g}_{ab}}={\Omega^2}(y){\eta_{ab}},{\bar{g}_{55}}=1,{\bar{g}_{5a}}={\bar{g}_{a5}}=0,\forall
     a$. Consider also the graviton perturbation ${\tilde{g}_{ab}}={\bar{g}_{ab}}+{h_{ab}},{\tilde{g}_{55}}={\bar{g}_{55}},{\tilde{g}_{5a}}={\bar{g}_{5a}}$ and the RS gauge
     ${\partial^a}{h_{ab}}=h=0$. Then
\beq
-{1\over{2}}{\bar{\nabla}^2}{h_{\mu\nu}}+{\bar{R}^{\alpha\;\beta}_{\,\mu\nu}}{h_{\alpha\beta}}+{\bar{G}^\beta_{(\mu}}{h_{\nu)\beta}}=-{\kappa_5^2}\delta{\mathcal{M}_{\mu\nu}}.
\eeq
Expanding we find
\beq
\left[-{1\over{2}}\left({\Omega^{-2}}{\Box_4}+{\partial_y^2}\right)+4{{{\partial_y^2}\Omega}\over{\Omega}}+4{{\left({{{\partial_y}\Omega}\over{\Omega}}\right)}^2}\right]{h_{ab}}=-{\kappa_5^2}\delta{\mathcal{M}_{ab}}
\eeq
The conformal 5D energy-momentum tensor $\tilde{\mathcal{T}_{\mu\nu}}$
is
\beq
{\tilde{\mathcal{T}}_{ab}}={{2{\Lambda_{\rm B}}{p_{\rm
        B}^5}}\over{\Omega^2}}{\tilde{g}_{ab}},\quad{\tilde{\mathcal{T}}_{a5}}={\tilde{\mathcal{T}}_{5a}}=0,\quad{\tilde{\mathcal{T}}_{55}}={{4{\Lambda_{\rm B}}{p_{\rm B}^5}}\over{\Omega^2}}{\tilde{g}_{55}}
\eeq
Then the energy-momentum tensor perturbation is 
\beq
\delta{\mathcal{M}_{ab}}={\Lambda_{\rm B}}\left(1-{{2{p_{\rm B}^5}}\over{\Omega^2}}\right){h_{ab}}+\left[\lambda\delta
      \left(y-{y_0}\right)+\lambda'\delta\left(y-{{y'}_0}\right)\right]{h_{ab}}
\eeq

\begin{figure}[H]
\center{\psfig{file=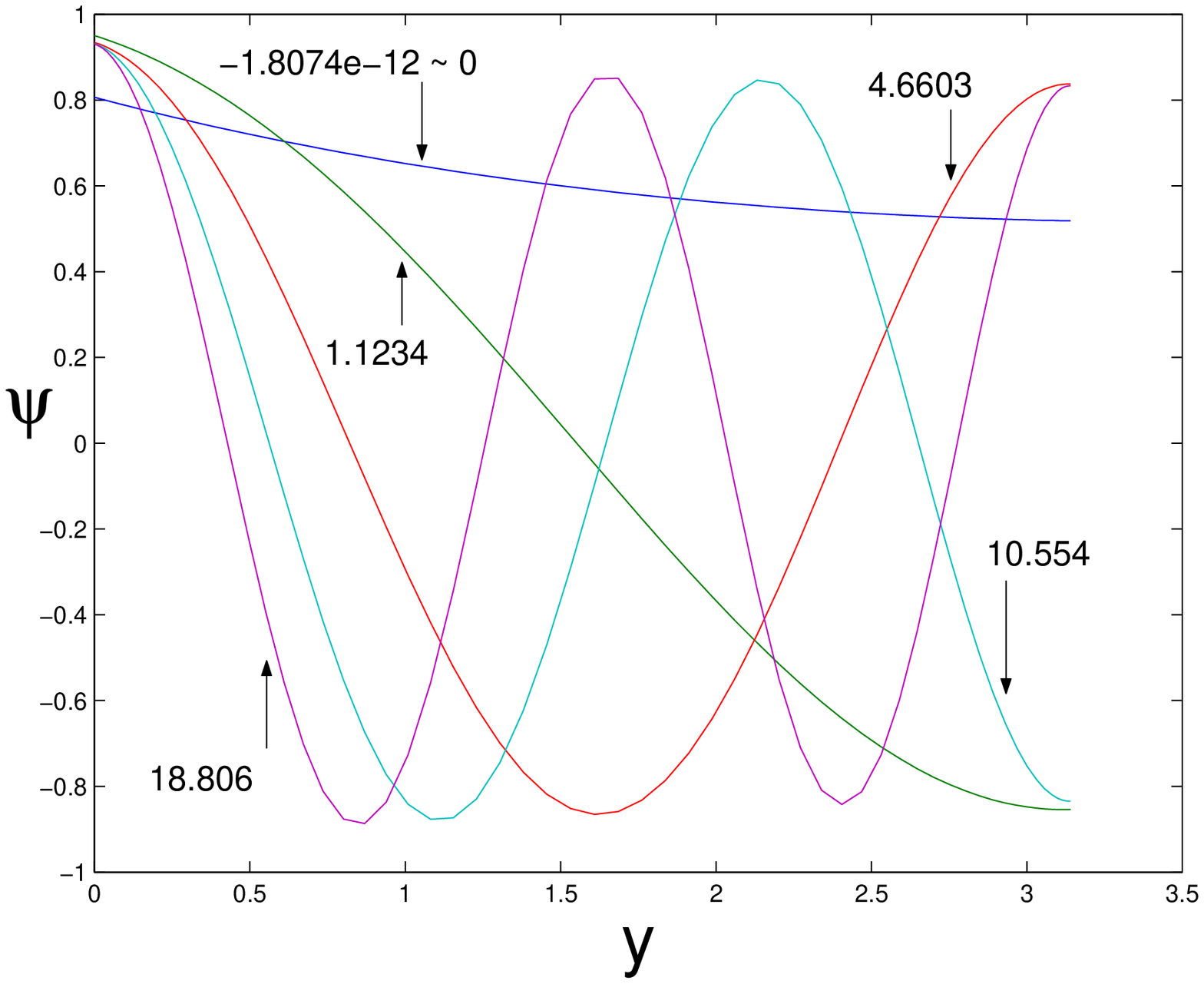,width=0.42\hsize}}
\vspace{-0.2cm}
\caption{\tiny Plot of graviton wave functions $\psi$ for $l/{r_{\rm c}}=5$ and
    ${p_{\rm B}^5}=0.25$. Shown are the massless graviton and the
    first four massive gravitons. The mass scale is set by
    ${r_{\rm c}}=100\mu$m$=1\times{{10^{3}}}{\mbox{eV}^{-1}}$.}
\label{fig2:matlab}
\end{figure}

Using the warp equations (\ref{rswf1}), (\ref{rswf2}) in the $y$
    coordinate and writting the graviton wave function as
    ${h_{ab}}={{\rm e}^{ip\cdot
    x}}{\psi_{ab}}(y)$ where ${p^2}=-{m^2}$ we obtain the following
    Sturm-Liouville problem
\[
 -{1\over{2}}{\partial_y^2}{\psi_{ab}}+{2\over{l^2}}\left[1-2{p_{\rm
       B}^5}{{\rm e}^{2|y|/l}}{{\left(1+{p_{\rm B}^5}{{\rm e}^{2|y|/l}}\right)}^{-2}}\right]{\psi_{ab}}
-{2\over{l}}{{1-{p_{\rm B}^5}}\over{1+{p_{\rm
          B}^5}}}\delta(y){\psi_{ab}}
\]
\beq
-{2\over{l}}{{1-{p_{\rm
          B}^5}{{\rm e}^{2\pi{r_{\rm c}}/l}}}\over{1+{p_{\rm
          B}^5}{{\rm e}^{2\pi{r_{\rm c}}/l}}}}\delta(y-\pi{r_{\rm c}}){\psi_{ab}}
={m^2}{{\rm e}^{2|y|/l}}{{\left(1+{p_{\rm B}^5}{{\rm e}^{2|y|/l}}\right)}^{-2}}{\psi_{ab}}
\eeq
The eigenvalues $m^2$ are positive or zero leading to $m\geq 0$. The
   set is infinite but discrete, ${m_i},i=0,1,\cdots +\infty$. There
   is just one massless graviton, ${m_0}=0$. It has a positive wave
   function $\psi_0$ in the fifth dimension. This wave function is localized near the
   positive tension branes. The massive gravitons have
    oscillating wave functions ${\psi_i}, i=1,\cdots+\infty$. Their masses and mass splittings decrease
   when $r_{\rm c}$ increases. For the solutions with a stabilized radion
   there is no hierarchy between graviton wave
   functions (see figure~\ref{fig2:matlab}) so that gravity is not
   strongly localized.

Like the radion the massive gravitons may be above the TeV scale. 
Their mass splittings have the same order
   of magnitude. The radion and the less massive graviton
   have similar masses. The massive gravitons may also be
   light. Newton's potential is
\beq
V(r)=-{{{G_4}mm'}\over{r}}\left(1+{{|{\psi_1}|^2}\over{|\psi_0|^2}}{{\rm
      e}^{-{m_1}r}}+\cdots\right)
\eeq
Since ${|{\psi_1}|^2}/{|\psi_0|^2}\sim 1$ we must have
${r_{\rm c}}<160\mu$m and ${m_1}>1\times{10^{-3}}$eV. 

\section{Conclusions}

In this paper we have analysed exact 5D solutions describing the
dynamics of $\rm{AdS}_5$ braneworlds when conformal fields of
weight -4 exist in the bulk. We have considered solutions for which gravity is localized near the
     brane and the dynamics on the
     brane is for example that of inhomogeneous dust, generalized dark radiation
     and homogeneous polytropic dark energy. We have seen that the
     radion may be stabilized using only the conformal bulk fields of
     weight -4 which generate the dynamics on the brane. This requires
     invariance of their equation of state under the radion
     perturbation, a stabilizing sector with a constant negative 5D
     pressure and new warp functions. We have also discussed graviton
     perturbations and determined their mass eigenvalues and wave
     functions from a Sturm-Liouville problem. Besides a massless graviton localized on the positive
     tension branes we have seen that this scenario involves an infinite discrete set of increasingly massive
     gravitons. We have shown that the new stabilizing warp functions
     are unable to generate a large mass hierarchy and that
     gravity is not strongly localized near the branes. Possibilities
     to overcome this problem are the introduction of supersymmetry
     \cite{susy} or of a non-conformal 5D scalar field \cite{GW} to
     stabilize the radion field.  This analysis is left for future
     research. Finally we have also shown that to satisfy the current 
observational constraints \cite{PDG} the radion and the 
     less massive graviton may either be
     heavier than $\sim 100$GeV corresponding to a small fifth dimension
     $\sim 1\times{10^{-15}}$mm or light with mass above $\sim
     1\times{10^{-3}}$ eV corresponding to a compactification scale of
     the order of $100\mu$m.
\vspace{1cm}

\leftline{\large \bf Acknowledgements}
\vspace{0.25cm}

We would like to thank Rui Guerra for his continual
  support. We also thank the financial support of {\it Funda\c {c}\~ao
  para a Ci\^encia e a Tecnologia} - FCT and {\it Fundo Social
  Europeu} - FSE under the contract
SFRH/BPD/7182/2001 ({\it III Quadro Comunit\'ario de Apoio}), of {\it
  Centro de Electr\'onica, Optoelectr\'onica e Telecomunica\c
  {c}\~oes} - CEOT,
  of {\it Centro
  Multidisciplinar de Astrof\'{\i}sica}-CENTRA and 
{\it Conselho de Reitores das Universidades
  Portuguesas} - CRUP with project {\it Ac\c {c}\~ao Integrada
  Luso-Espanhola} E-126/04.

\end{document}